\documentclass[fleqn,usenatbib,letters]{mnras}

\usepackage{newtxtext,newtxmath}
\usepackage[T1]{fontenc}

\DeclareRobustCommand{\VAN}[3]{#2}
\let\VANthebibliography\thebibliography
\def\thebibliography{\DeclareRobustCommand{\VAN}[3]{##3}\VANthebibliography}

\usepackage{graphicx}	% Including figure files
\usepackage{amsmath}	% Advanced maths commands

\newcommand{\msol}{\mbox{$M_\odot$}}

\newcommand{\HI}{H\,{\sc i} }
\newcommand{\hi}{\ifmmode{\rm HI}\else{H\/{\sc i}}\fi} 
\newcommand {\kms}{\ifmmode{\rm km \, s^{-1}}\else{$\rm km \, s^{-1}$}\fi} 

\newcommand{\msold}{\mbox{$M_\odot$}}

\newcommand{\Sci}{Science}

\title[]{Unveiling a Thin Filament of the Cosmic Web in the Ursa Major Supergroup}

\author[J. L. Xu et al.]{Jin-Long Xu$^{1,2}$\thanks{E-mail: xujl@bao.ac.cn}, Ming Zhu$^{1,2}$, Peng Jiang$^{1,2}$, Nai-Ping Yu$^{1,2}$, Chuan-Peng Zhang$^{1,2}$, \newauthor{Xiao-Lan Liu$^{1,2}$, Mei Ai$^{1,2}$, Yin-Jie Jing$^{1}$, Jie Wang$^{1}$}
\\\\
$^{1}$National Astronomical Observatories, Chinese Academy of Sciences, Beijing 100101, People's Republic of China\\
$^{2}$Guizhou Radio Astronomical Observatory, Guizhou University, Guiyang 550000, People's Republic of China}

\date{Accepted 2025 -- --. Received 2025 -- --; in original form 2025 -- --}

\begin{document}
\label{firstpage}
\pagerange{\pageref{firstpage}--\pageref{lastpage}}
\maketitle

\begin{abstract}
Filaments are crucial components of the cosmic web, representing the extensive and aligned distributions of galaxies and gas.  Using the Five-hundred-meter Aperture Spherical radio Telescope (FAST), we report the detection of a filament in the Ursa Major supergroup using atomic-hydrogen (\hi) observations. This filament consists of sixteen various types of galaxies and five starless gas clumps, spanning a length of approximately 0.9 Mpc. Notably, it is extremely thin, with a thickness comparable to the diameter of a galaxy.  We observed  a  galaxy-filament spin alignment and  a velocity gradient within the filament. These findings strongly suggest  a cold accretion flow along the filament, potentially contributing to the formation and growth of the galaxies. The thin filament, as a small group, is likely to be merged into the Ursa Major supergroup in the context of hierarchical structure formation.
\end{abstract}


%%%%%%%%%%%%%%%%%%%%%%%%%%%%%%%%%%%%%%%%%%%%%%%%%%

%%%%%%%%%%%%%%%%% BODY OF PAPER %%%%%%%%%%%%%%%%%%

\section{Introduction}
\label{sec:intro}

Cosmological simulations of structure formation within the cold dark matter framework predict that the cosmic web resembles a spiderweb with filaments acting as threads, separated by vast, nearly empty regions called voids \citep{1978MNRAS.185..357J,2014MNRAS.441.2923C}. These filaments consist of dark matter, gas, and galaxies, and may represent up to half of the baryonic mass budget of the Universe. Streams of cold gas falling along these filaments are expected to supply most of the gas necessary for the galaxy growth \citep{2019Sci...366...97U}. Impressively, the filaments are predicted to exhibit hierarchical structures \citep{2010MNRAS.408.2163A,2012MNRAS.422...25S,2020Natur.585...39W}. Galaxy clusters and groups form at the intersections of the large-scale filaments \citep{2005MNRAS.359..272C,2012Natur.487..202D}. Cosmological simulations also suggest the presence of smaller-scale filaments consisting of dwarf galaxies interspersed among normal galaxies \citep{1996Natur.380..603B}.  Observational evidence for diffuse substructures remains scarce.  Detecting such small-scale filaments is critical to test predictions of structure formation and to better determine the filamentary environment that affects galaxy formation and evolution.  

Due to the faintness of any gas emissions in the filaments \citep{2017MNRAS.468..857K,2019MNRAS.490.1415K,2019Sci...366...97U}, current detection relies mainly on the characteristic distribution of bright galaxies identified through galaxy redshift surveys \citep{2014MNRAS.438.3465T,2022A&A...659A.166C,2022ApJS..259...43C}. Since the limitations of current optical survey in detecting brightness, it is not yet possible to obtain spectroscopic information of weak dwarf galaxies.
This approach through bright galaxies has predominantly led to the discovery of large-scale, thicker filaments, typically with a physical radius of around 1 to 2 megaparsec (Mpc) \citep{2018MNRAS.473.1195L}, and further impede the detection of the smaller filaments.  The observation of atomic-hydrogen (\hi) is more likely to determine the redshift of faint galaxies \citep{2025ApJ...982L..36X}. Moreover, given that \HI gas extends well beyond the optical emission in gas-rich dwarf galaxies, the \HI observations may be more effective for identifying the smaller galaxy filaments. A currently found smaller filament from the \HI observation is composed of three interacting galaxies in a void \citep{2013AJ....145..120B}.

Using the FAST, we set out to carry out an extragalactic \HI survey (FASHI) \citep{2024SCPMA..6719511Z}, which aims to search for dark and faint galaxies, utilizing the high sensitivity of FAST combined with high velocity resolution.  In this letter, we report the detection of a thin  filament in the nearby Ursa Major supergroup based on the \HI observations obtained using the FAST.

\section{Observations and Parameter calculation}
\label{sec:obs}
\subsection{Observations and data processing of FAST}
Based on the FASHI survey, we have found a group of galaxies with a nearly linear distribution. To detect the fainter galaxies in this group, multiple observations from the FAST \citep{2019SCPMA..6259502J,2020RAA....20...64J} were performed  using the Multi-beam on-the-fly (OTF) mode from September to December 2023. This mode is designed to map the sky with 19 beams simultaneously and has a similar scanning trajectory. The scan velocity was set to 15$^{\prime\prime}$ s$^{-1}$ and the integration time to 1 second per spectrum. It formally works in the frequency range from 1050 MHz to 1450 MHz.  We utilized the digital Spec(W) backend, which has a bandwidth of 500 MHz and 64k channels. This results in a frequency resolution of 7.629 kHz, corresponding to a velocity resolution of 1.6 \kms. To calibrate intensity to antenna temperature ($T_{\rm A}$), a noise signal with an amplitude of 10 K was injected over a period of 32 seconds. The half-power beam width (HPBW) for each beam is $\sim$2.9$^{\prime}$ at 1.4 GHz. The pointing accuracy of the telescope was better than 10$^{\prime\prime}$.  System temperature during the observations was around 22 K. The detailed data reduction is similar to \citep{2021ApJ...922...53X}. Finally, the calibrated data were converted to the standard cube data, with a pixel size of $1.0^{\prime}\times1.0^{\prime}$. A gain $T_{\rm A}/\it S_{v}$ was measured to be about 16 K Jy$^{-1}$. While a measured relevant main beam gain $T_{\rm B}/\it S_{v}$ is about 20 K Jy$^{-1}$ at 1.4 GHz for each beam, where $T_{\rm B}$ is the brightness temperature.  To construct a highly sensitivity \HI image, the velocity resolution of the FAST data is smoothed to 6.4 \kms. The mean noise rms in the observed image is $\sim$0.46 mJy beam$^{-1}$ in the observed region.

\subsection{Parameter calculation}
The baryonic mass of galaxy can be estimated by $M_{\rm bar}=M_{\star}+M_{\rm gas}$, where $M_{\star}$ is the total stellar mass. The gas content of galaxies is primarily composed of \HI and helium. To account for the contribution of Helium, a factor of 1.33 is included assuming the same helium-to-\HI ratio.  The total gas mass of galaxies can be derived by $M_{\rm gas} = 1.33\times M_{\rm HI}$, where  $M_{\rm HI}$ is the \HI gas mass determined by $M_{\rm HI} =  2.36\times10^{5}D^{2}S_{a}$, where $D$ is the adopted distance to each galaxy and $S_{a}$ is flux of each galaxy. Here, we use the distance of the Ursa Major supergroup as the distance of each galaxy.   

To estimate the total stellar mass of each galaxy in the filament, we used $g$-band and $r$-band data from the Sloan Digital Sky Survey (SDSS DR12)\citep{2015ApJS..219...12A}. The stellar mass of galaxy can be given by log$(M_{\star}/L_{g})=$-0.745+1.616$(m_{g}-m_{r})$ \citep{2017ApJS..233...13Z}, where $L_{g}$ is the stellar luminosity in $g$-band, which can be determined by $L_{g}$=$D^{2}10^{10-0.4(m_{g}-M_{g}^{0})}$, where $M_{g}^{0}$ is the absolute solar g-band magnitude, which is assumed to be 5.03 mag \citep{2018ApJS..236...47W}.  The $m_{g}$, $m_{r}$ and position-angle (PA) for each galaxy are obtained using the software SEXtractor \citep{1996A&AS..117..393B}. The PA is counted from the North (PA=0$^{\circ}$) toward East between 0$^{\circ}$ and 180$^{\circ}$. 

Assuming that the filament is an axisymmetric and isothermal cylinder,  its total mass (baryonic and dark matter) can be given by \citep{1997ApJ...475..421E}:
\begin{equation}
      M_{\rm tot}=2L\sigma_{\perp}^{2}/G
\end{equation}
where $G$ is the gravitational constant, $L$ is the length of the filament in Mpc, and $\sigma_{\perp}$ is the transverse velocity dispersion, which can be calculated using the line-of-sight velocities of the galaxies and gas clumps in the filament. Here $\sigma_{\perp}$  can be determined by:
\begin{equation}
    \sigma_{\perp}=\sqrt{\langle\upsilon^{2}\rangle-\langle\upsilon\rangle^{2}}
\end{equation}
where $\upsilon$ is the array of line-of-sight velocities for the filament.

\begin{figure*}
\centering
\includegraphics[width=0.95\textwidth]{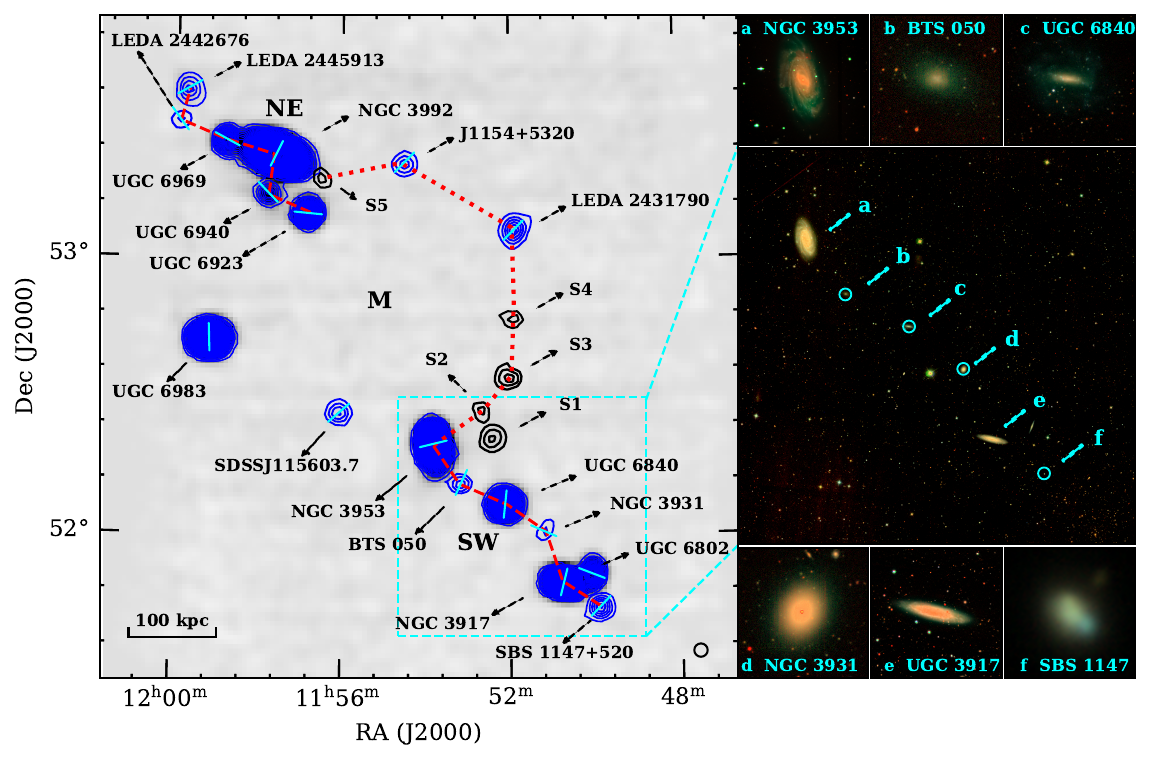}
\vspace{-3mm}
\caption{Overview of the galaxy filament.  Left, \HI column-density map shown in the grey scale and blue contours for each galaxy. The contours begin at 3.2$\times$10$^{17}$ cm$^{-2}$ (3$\sigma$) in steps of 3.8$\times$10$^{17}$ cm$^{-2}$. The black contours indicate several gas clumps. The red dashed and dotted lines are used to link two adjacent galaxies. The cyan lines on each galaxy indicate the direction of spin.  The FAST beam in a black circle is shown in the bottom-right corner. Right, the SDSS RGB (z, r,  and g bands) composite image of a small region shown by a cyan dashed box in the \HI column-density map. }
\label{fig:Filament_all}
\end{figure*}

\section{Results}
\label{sect:results}
Using the FAST \HI observations, we identified a group of galaxies with a nearly linear distribution extending from northeast to southwest. To identify galaxies in our observing region, we used the Source Finding Application (SoFiA2) in the \HI data cube \citep{2021MNRAS.506.3962W}. For each identified source, we also obtained several parameters, including the coordinate position and radial velocity. Using the NASA/IPAC Extragalactic Database (NED), we searched for the optical counterparts of these sources. Of these sources, 16 are known galaxies and one newly identified galaxy (J1154+5320), while the remaining are gas clumps without optical counterparts as analyzed in the Sloan Digital Sky Survey (SDSS) and the Dark Energy Spectroscopic Instrument (DESI) images. These starless gas clumps are likely to be dark galaxies \citep{2021A&A...649L..14R,2025SciA...11S4057L} or residual gases. Due to the resolution limitations of the FAST, we are still unable to decipher the properties of these clumps. We will apply for a higher resolution interferometric array to observe these clumps. Moreover, with this high detection sensitivity, we detected \HI gas associated with only two lenticular (S0) galaxies in this group. Figure  \ref{fig:Filament_all} illustrates that the \HI emission from these galaxies and gas clumps assembles a thin filament of galaxies, with a thickness comparable to the diameter of a galaxy.  The observed parameters for the galaxies and the clumps  are summarized in Table~\ref{tab:Obs}. 

\begin{table*} 
     \centering
      \caption{Measured and derived properties of galaxies and gas clumps in the filament.  We list: system velocity ($V_{\rm sys}$); position angle ($PA$); apparent magnitude ($m_{g}$, $m_{r}$), which has been corrected for extinction;  steller mass ($M_{\rm \star}$);  \HI gas mass ($M_{\rm HI}$). }
      \vspace{-8pt}
      \setlength{\tabcolsep}{9pt}
      \label{tab:Obs}
      \begin{tabular}{lccccccccccc}
\noalign{\vspace{5pt}}\hline\hline\noalign{\vspace{5pt}}
Name & Type & RA & Dec. & $V_{\rm sys}$ & $PA$ &$m_{g}$ & $m_{r}$ & log$M_{\rm \star}$ & log$M_{\rm HI}$ \\
 &   &   & [\kms] &[\kms] & [$\circ$]  &[mag] & [mag] & [$\msol$]&[$\msol$]\\
\noalign{\vspace{5pt}}\hline\noalign{\vspace{5pt}}
SBS 1147+520 & Irr &11$^{\rm h}$49$^{\rm m}$54.4$^{\rm s}$ & 51$^{\rm \circ}$44$^{\rm \prime}$11.0$^{\rm \prime\prime}$ &  943.0$^{+0.7}_{-0.7}$  & 28.3$^{+2.1}_{-2.1}$  & 16.6$^{+0.1}_{-0.1}$ & 16.4$^{+0.1}_{-0.1}$ &7.7$^{+0.1}_{-0.1}$ &7.4$^{+0.1}_{-0.1}$   \\

NGC 3917 & S &11$^{\rm h}$50$^{\rm m}$45.4$^{\rm s}$ & 51$^{\rm \circ}$49$^{\rm \prime}$28.9$^{\rm \prime\prime}$ &  957.5$^{+0.4}_{-0.4}$  & 76.4$^{+1.0}_{-1.0}$  & 12.2$^{+0.1}_{-0.1}$ & 11.6$^{+0.1}_{-0.1}$ & 9.7$^{+0.1}_{-0.1}$ & 9.3$^{+0.1}_{-0.1}$   \\

UGC 6802 & S &11$^{\rm h}$50$^{\rm m}$06.5$^{\rm s}$ & 51$^{\rm \circ}$51$^{\rm \prime}$21.3$^{\rm \prime\prime}$ &  1250.8$^{+0.8}_{-0.8}$  &159.1$^{+0.5}_{-0.5}$  & 14.9$^{+0.1}_{-0.1}$  & 14.6$^{+0.1}_{-0.1}$ &8.5$^{+0.1}_{-0.1}$ &8.7$^{+0.1}_{-0.1}$   \\

NGC 3931 & S0 &11$^{\rm h}$51$^{\rm m}$13.5$^{\rm s}$ & 52$^{\rm \circ}$00$^{\rm \prime}$03.1$^{\rm \prime\prime}$ &  954.0$^{+2.1}_{-2.1}$  & 156.9$^{+2.4}_{-2.4}$  & 13.7$^{+0.1}_{-0.1}$ & 13.1$^{+0.1}_{-0.1}$ &9.1$^{+0.1}_{-0.1}$ & 6.5$^{+0.1}_{-0.1}$\\

UGC 6840 & SB &11$^{\rm h}$52$^{\rm m}$07.0$^{\rm s}$ & 52$^{\rm \circ}$06$^{\rm \prime}$28.8$^{\rm \prime\prime}$ &  1017.7$^{+0.4}_{-0.4}$  &71.7$^{+4.2}_{-4.2}$  & 14.1$^{+0.1}_{-0.1}$  & 13.8$^{+0.1}_{-0.1}$ &8.7$^{+0.1}_{-0.1}$ &9.1$^{+0.1}_{-0.1}$   \\

BTS 050 & dS0 &11$^{\rm h}$53$^{\rm m}$09.3$^{\rm s}$ & 52$^{\rm \circ}$11$^{\rm \prime}$22.4$^{\rm \prime\prime}$ &  1220.6$^{+2.0}_{-2.0}$  & 65.57$^{+1.7}_{-1.7}$  & 16.1$^{+0.1}_{-0.1}$  & 15.2$^{+0.1}_{-0.1}$ &8.3$^{+0.1}_{-0.1}$ &6.8$^{+0.2}_{-0.2}$   \\

NGC 3953 &SB& 11$^{\rm h}$53$^{\rm m}$48.9$^{\rm s}$ & 52$^{\rm \circ}$19$^{\rm \prime}$36.4$^{\rm \prime\prime}$ &  1043.8$^{+4.1}_{-4.1}$  & 13.7$^{+1.3}_{-1.3}$  & 10.5$^{+0.1}_{-0.1}$  & 9.7$^{+0.1}_{-0.1}$ &10.5$^{+0.1}_{-0.1}$ &9.5$^{+0.1}_{-0.1}$   \\

SDSSJ115603.7 &Irr& 11$^{\rm h}$56$^{\rm m}$03.4$^{\rm s}$ & 52$^{\rm \circ}$26$^{\rm \prime}$17.1$^{\rm \prime\prime}$ &  879.0$^{+4.0}_{-4.0}$  & 40.8$^{+3.2}_{-3.2}$  & 17.5$^{+0.1}_{-0.1}$ &16.9$^{+0.1}_{-0.1}$ & 7.6$^{+0.1}_{-0.1}$ &7.1$^{+0.1}_{-0.1}$   \\

UGC 6983 &SB & 11$^{\rm h}$59$^{\rm m}$09.3$^{\rm s}$ & 52$^{\rm \circ}$42$^{\rm \prime}$27.0$^{\rm \prime\prime}$ &  1083.8$^{+0.4}_{-0.4}$  &91.3$^{+1.5}_{-1.5}$  & 12.8$^{+0.1}_{-0.1}$ & 12.6$^{+0.1}_{-0.1}$ & 9.2$^{+0.1}_{-0.1}$ &9.5$^{+0.1}_{-0.1}$  \\

LEDA 2431790 &Irr& 11$^{\rm h}$51$^{\rm m}$53.7$^{\rm s}$ & 53$^{\rm \circ}$05$^{\rm \prime}$58.3$^{\rm \prime\prime}$ &  1123.0$^{+1.1}_{-1.1}$  & 45.6$^{+3.7}_{-3.7}$  & 16.7$^{+0.1}_{-0.1}$  & 16.2$^{+0.1}_{-0.1}$ &7.8$^{+0.1}_{-0.1}$ &7.3$^{+0.1}_{-0.1}$   \\

UGC 6923 &S& 11$^{\rm h}$56$^{\rm m}$49.4$^{\rm s}$ & 53$^{\rm \circ}$09$^{\rm \prime}$37.3$^{\rm \prime\prime}$ &  1069.7$^{+0.3}_{-0.3}$  &174.2$^{+2.2}_{-2.2}$  & 13.7$^{+0.1}_{-0.1}$   & 13.3$^{+0.1}_{-0.1}$ &9.0$^{+0.1}_{-0.1}$ &8.9$^{+0.1}_{-0.1}$  \\

UGC 6940 &S& 11$^{\rm h}$57$^{\rm m}$47.6$^{\rm s}$ & 53$^{\rm \circ}$14$^{\rm \prime}$04.1$^{\rm \prime\prime}$ &  1114.2$^{+0.5}_{-0.5}$  & 133.5$^{+0.8}_{-0.8}$  & 16.1$^{+0.1}_{-0.1}$ & 15.9$^{+0.1}_{-0.1}$ & 7.9$^{+0.1}_{-0.1}$ &8.2$^{+0.1}_{-0.1}$  \\

J1154+5320 &Irr& 11$^{\rm h}$54$^{\rm m}$30.6$^{\rm s}$ & 53$^{\rm \circ}$20$^{\rm \prime}$46.9$^{\rm \prime\prime}$ &  1168.0$^{+1.4}_{-1.4}$  & 45.4$^{+2.2}_{-2.2}$  & 18.9$^{+0.5}_{-0.5}$ & 18.3$^{+0.6}_{-0.6}$ & 7.0$^{+0.1}_{-0.1}$ &6.9$^{+0.1}_{-0.1}$   \\

NGC 3992 &SB& 11$^{\rm h}$57$^{\rm m}$36.0$^{\rm s}$ & 53$^{\rm \circ}$22$^{\rm \prime}$29.0$^{\rm \prime\prime}$ &  1059.1$^{+0.5}_{-0.5}$  & 63.5$^{+1.6}_{-1.6}$  & 10.4$^{+0.1}_{-0.1}$  & 9.7$^{+0.1}_{-0.1}$ &10.4$^{+0.1}_{-0.1}$ &9.7$^{+0.1}_{-0.1}$   \\

UGC 6969 &S& 11$^{\rm h}$58$^{\rm m}$47.6$^{\rm s}$ & 53$^{\rm \circ}$25$^{\rm \prime}$29.1$^{\rm \prime\prime}$ & 1117.2$^{+1.0}_{-1.0}$  &151.2$^{+2.0}_{-2.0}$  & 14.8$^{+0.1}_{-0.1}$ & 14.6$^{+0.1}_{-0.1}$ &  8.4$^{+0.1}_{-0.1}$ & 8.6$^{+0.1}_{-0.1}$     \\

LEDA 2442676  & Irr & 11$^{\rm h}$59$^{\rm m}$56.2$^{\rm s}$ & 53$^{\rm \circ}$29$^{\rm \prime}$44.8$^{\rm \prime\prime}$ &  1057.8$^{+2.0}_{-2.0}$   & 125.4$^{+2.2}_{-2.2}$  & 16.7$^{+0.1}_{-0.1}$ & 16.3$^{+0.1}_{-0.1}$ & 7.8$^{+0.1}_{-0.1}$ & 6.9$^{+0.2}_{-0.2}$   \\

LEDA 2445913 &Irr& 11$^{\rm h}$59$^{\rm m}$43.2$^{\rm s}$ & 53$^{\rm \circ}$36$^{\rm \prime}$38.8$^{\rm \prime\prime}$ &  1139.7$^{+0.8}_{-0.8}$  &28.6$^{+2.8}_{-2.8}$  & 17.5$^{+0.2}_{-0.2}$ & 17.3$^{+0.2}_{-0.1}$ &7.3$^{+0.1}_{-0.1}$ &7.4$^{+0.1}_{-0.1}$   \\

S1 & & 11$^{\rm h}$52$^{\rm m}$25.3$^{\rm s}$ & 52$^{\rm \circ}$20$^{\rm \prime}$46.5$^{\rm \prime\prime}$ &  1119.2$^{+1.1}_{-1.1}$  & -- & -- & -- & -- &7.0$^{+0.1}_{-0.1}$   \\

S2 & & 11$^{\rm h}$52$^{\rm m}$39.8$^{\rm s}$ & 52$^{\rm \circ}$26$^{\rm \prime}$53.2$^{\rm \prime\prime}$ &  1012.4$^{+4.3}_{-4.3}$ & -- & -- & -- & --  &6.6$^{+0.2}_{-0.2}$   \\

S3 & & 11$^{\rm h}$52$^{\rm m}$01.0$^{\rm s}$ & 52$^{\rm \circ}$33$^{\rm \prime}$56.3$^{\rm \prime\prime}$ &  1050.2$^{+2.9}_{-2.9}$ & -- & -- & -- & -- &6.9$^{+0.1}_{-0.1}$   \\

S4 && 11$^{\rm h}$51$^{\rm m}$54.5$^{\rm s}$ & 52$^{\rm \circ}$46$^{\rm \prime}$46.7$^{\rm \prime\prime}$ &  1071.8$^{+2.1}_{-2.1}$ & -- & -- & -- & --  & 6.5$^{+0.2}_{-0.2}$   \\

S5 && 11$^{\rm h}$56$^{\rm m}$30.4$^{\rm s}$ & 53$^{\rm \circ}$17$^{\rm \prime}$18.7$^{\rm \prime\prime}$ &  1168.8$^{+1.9}_{-1.9}$ & -- & -- & -- & -- &6.8$^{+0.2}_{-0.2}$   \\
\noalign{\vspace{5pt}}\hline\hline\noalign{\vspace{5pt}}
\end{tabular}
\\Note: Irr--dwarf irregular galaxy;  S--spiral galaxy; SB--barred spiral galaxy; S0--lenticular galaxy; dS0--dwarf lenticular galaxy.
\end{table*}

\begin{figure}
\centering
\includegraphics[width=0.35\textwidth]{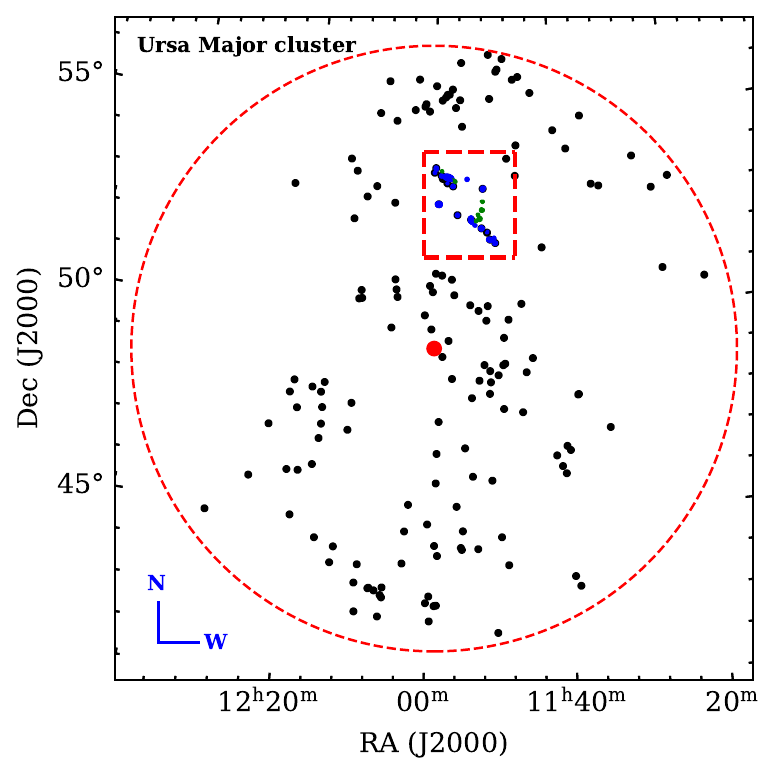}
\vspace{-3mm}
\caption{Overview of the Ursa Major supergroup from the optical observations \citep{2013MNRAS.429.2264K,2014MNRAS.445..630P}.  The center and size of the supergroup are indicated by the red solid and dashed circles, respectively. The identified filament is marked with a red dashed box. The black dots represent the positions of the member galaxies in the supergroup, while the blue and green dots indicate the member galaxies and \HI gas clumps in the filament.  N, north; W, west.}
\label{fig:cluster}
\end{figure}

The identified filament, composed of seventeen galaxies and five gas clumps, is located at about 1.1 Mpc north of the center of the Ursa Major cluster, as shown in Fig. \ref{fig:cluster}. This galaxy cluster, with a distance of 17.4$\pm$0.3 Mpc \citep{2008ApJ...676..184T,2012ApJ...749...78T} obtained from a newly available Cepheid distances to a local Tully-Fisher calibrator,  has a radius of about 7.5$^{\circ}$ \citep{1996AJ....112.2471T}. With different optical observations \citep{2013MNRAS.429.2264K,2014MNRAS.445..630P}, a total of 166 galaxies were detected in this galaxy cluster.  Recent studies show that the Ursa Major cluster does not meet the criteria for a galaxy cluster. However, it is more complex than a typical group, so the Ursa Major cluster is classified as a supergroup \citep{2016PASA...33...38W,2025ApJS..278...37Y}. Previous studies have shown that this supergroup has heliocentric velocities ranging from 700.0 to 1210.0 \kms and a low velocity dispersion of 148.0 \kms \citep{1996AJ....112.2471T}. Our observations indicate that the filament spans a velocity range between 876.8 and 1250.8 \kms and has a velocity dispersion  of 150.0 \kms. These velocities are consistent with those of the Ursa Major supergroup, suggesting that the identified filament is part of this nearby supergroup. Furthermore, due to the low velocity scatter for a linear distribution of galaxies, we have ruled out the possibility that this filament is an artifact of a chance orientation effect within the supergroup.

\begin{figure}
\centering
\includegraphics[width=0.43\textwidth]{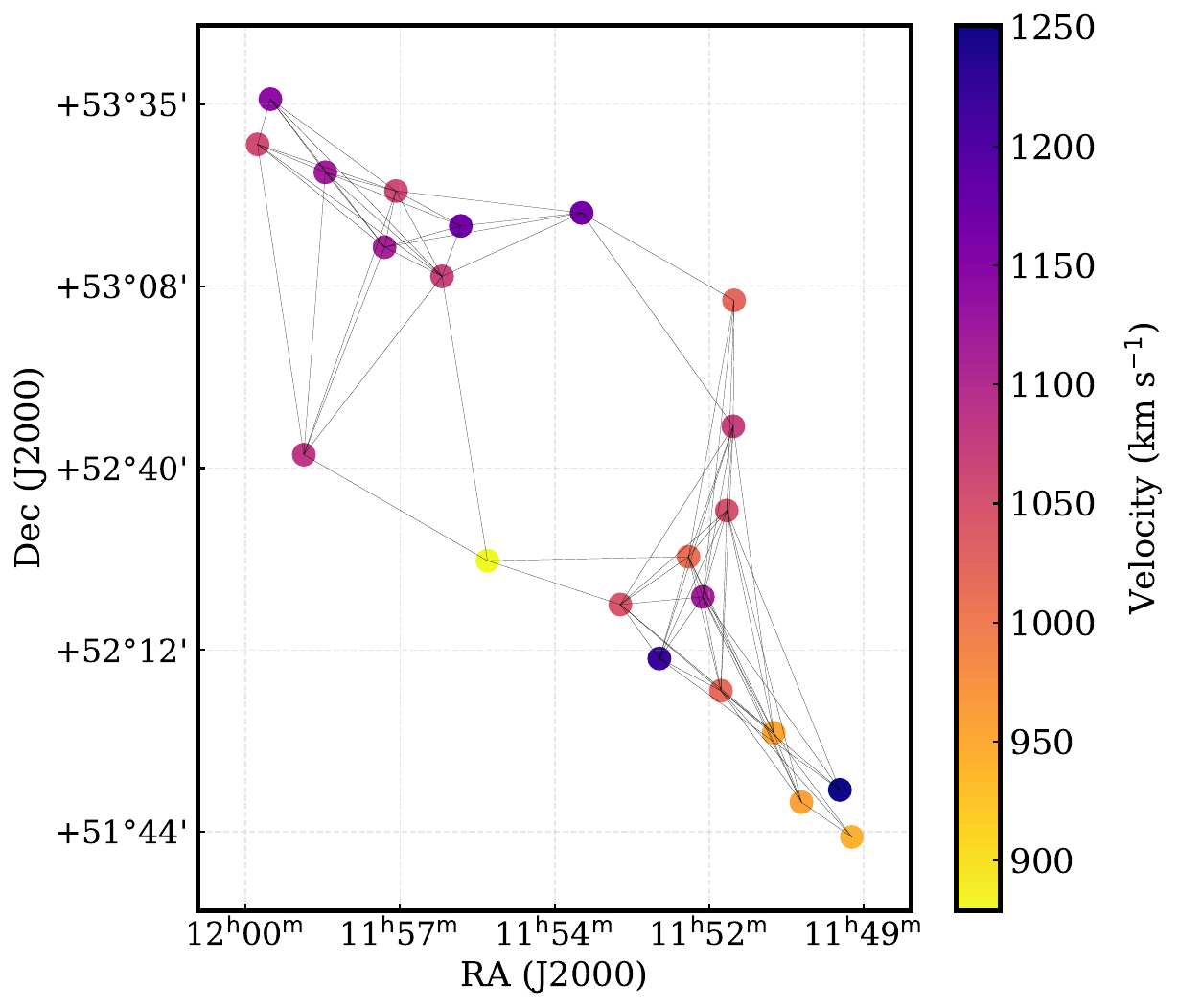}
\vspace{-3mm}
\caption{Detected filamentary pattern in a 22-source volume within a pattern of galaxies and clumps (points). The initial linking length values. The short black line represents two galaxies that comply with the FoF algorithm.}
\label{fig:filament}
\end{figure}

To check whether this filament is consistent in overall dynamics, we used the friends-of-friends (FoF) algorithm \citep{2010A&A...514A.102T,2012A&A...540A.106T,2014MNRAS.438.3465T}. The initial linking length values are 250 \kms for the radial length and 0.25 Mpc for the transversal length \citep{2010A&A...514A.102T}. Figure \ref{fig:filament} shows the detected filamentary pattern in a 22-source volume within a pattern of galaxies and clumps (points). From the Fig. \ref{fig:filament}, we see that the statistical test also supports our discovery that these sources form a coherent filament. Following this discovery, we sought to investigate whether the sub-sample displayed any kinematic coherence in the filament. For discussion, we will divide the filament into three sections: northeast (NE), middle (M), and southwest (SW). The velocity difference between the member galaxy UGC 6802, located in the SW section of this filament, and its three neighboring galaxies is approximately 300 \kms. This substantial velocity difference suggests that UGC 6802 may be a background galaxy relative to the filament. The SW section is particularly notable for its geometric configuration. The six galaxies in the SW section assemble a smaller filament, which appears to exhibit a fluctuating structure, as shown in Fig. \ref{fig:Filament_all}. This SW filament connects S0 galaxies and a dwarf galaxy through massive spiral galaxies. Such a structure is analogous to a filamentary cluster-cluster observed in cosmological simulations, while superclusters are bridges \citep{1996Natur.380..603B}.

\begin{figure}
\centering
\includegraphics[width=0.47\textwidth]{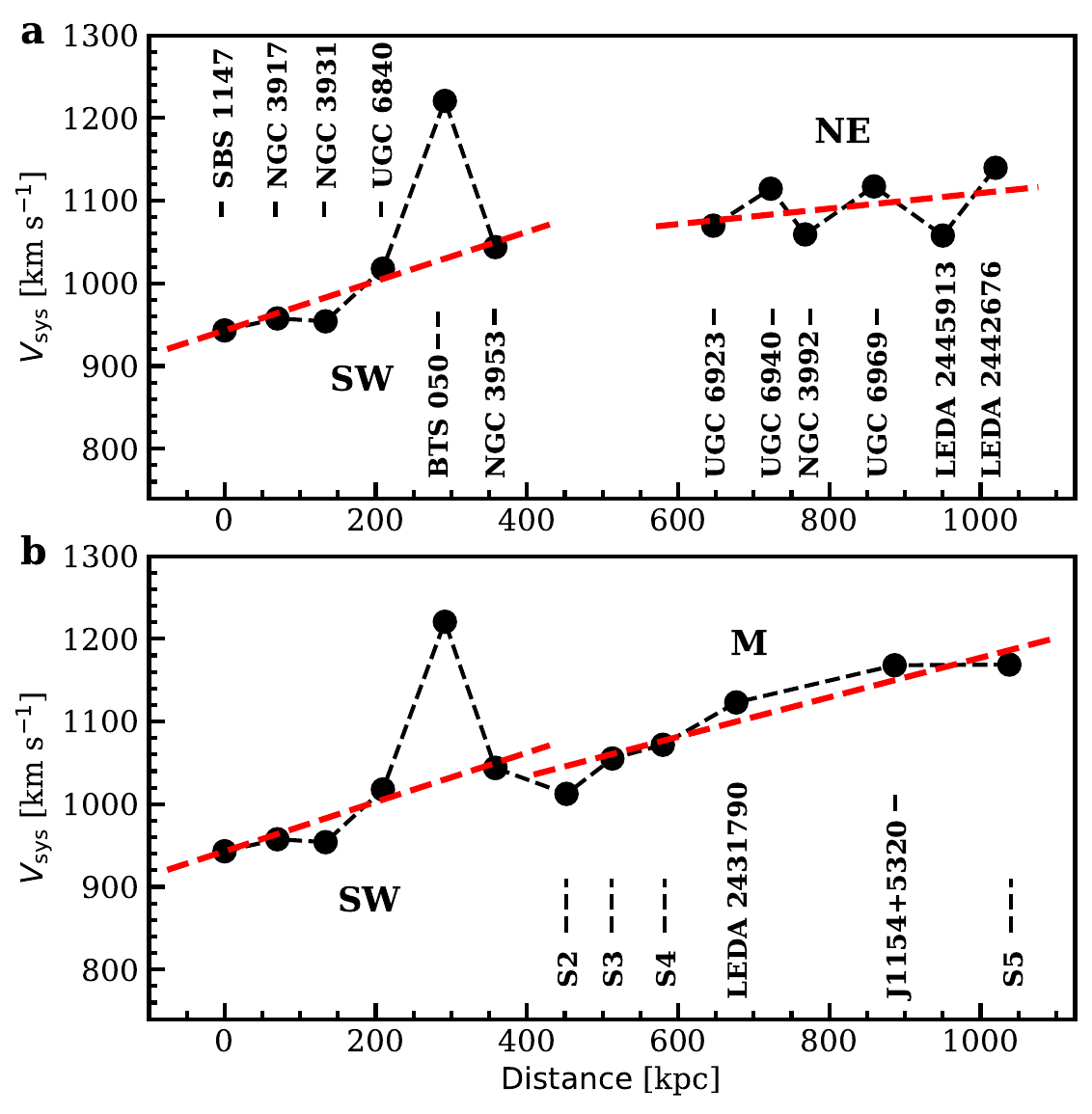}
\vspace{-3mm}
\caption{Velocity--distance plot. Each black dot represents a galaxy. The starting dot is the galaxy SBS 1147 located at the bottom of the galaxy filament. Each two galaxies are linked by a black dashed line. The red dashed lines are the fitted curves for the galaxies in the SW, NE, and M smaller filaments. a, the small filaments in the SW and M sections of the galaxy filament.  b, the small filaments in the SW and NE sections of the galaxy filament.}
\label{fig:Filament_DV}
\end{figure}

In Fig. \ref{fig:Filament_all}, the right panel shows that the galaxies in the SW filament are arranged in a manner reminiscent of a pearl necklace, with evenly spaced intervals of approximately 71.6 kpc between adjacent galaxies. We constructed a velocity-distance map of the galaxies along the spine of the SW filament, as shown in Fig. \ref{fig:Filament_DV}. This analysis revealed a velocity gradient within the SW filament, with a linear fit revealing a gradient of 348$\pm$44 \kms Mpc$^{-1}$. Notably, the velocity of BTS 050 deviates about 180 \kms from the fit, potentially indicating the unique formation process of this dwarf S0 galaxy. Detections of a velocity gradient implies that the SW filament exhibits characteristics of a coherent structure.

In the NE section, six galaxies and a gas clump assemble a smaller filament that extends toward the spine of the SW filament.  A velocity-distance map (Fig. \ref{fig:Filament_DV}) along the spine of the NE filament reveals a slight velocity gradient. In the M section, the northwest part consists of a smaller filament composed of dwarf galaxies and gas clumps, as shown by a red dotted line in Fig. \ref{fig:Filament_all}. We obtained that the M  and NE filaments have velocity gradients of 235$\pm$71 \kms Mpc$^{-1}$ and 94$\pm$12 \kms Mpc$^{-1}$ from Fig. \ref{fig:Filament_DV}, respectively. Hence, both the NE and M filaments are coherent structures. From an overall perspective, the SW filament is coherent with the NE filament in terms of velocity and space, while the M filament serves as the bridge. The M filament seems to be a filament of dwarf galaxies as predicted by numerical simulations\citep{1996Natur.380..603B}.

To investigate whether the galaxy filament exhibits additional kinematic coherences, we analyzed the correlation between the spin axes of the galaxies and the orientation of their host filament. The spin axes of each galaxy are indicated by a cyan line in Fig. \ref{fig:Filament_all}.  We considered a galaxy to have a spin axis perpendicular to the filament if the PA difference from the filament was less than 45$^{\circ}$. The PA of the filament was measured to be approximately 46.3$^{\circ}$. We found that, except for four galaxies adjacent to NGC 3992 and the S0 galaxy NGC 3931, all galaxies had spin axes perpendicular to the filament. Previous observations have shown that the spins of S0 galaxies differ from those of other galaxies in filaments  \citep{2013MNRAS.428.1827T,2013ApJ...779..160Z,2021MNRAS.504.4626K}. Additionally, satellite galaxies have been observed to align with galaxy filaments \citep{2015MNRAS.450.2727T} and orient toward the host galaxy due to tidal interaction \citep{2006ApJ...644L..25A,2007ApJ...671.1135K}. The four galaxies with inconsistent spin orientations were found to be pointing toward NGC 3992, suggesting that they may be satellite galaxies of this massive galaxy. Additionally, the spin axis of UGC 6802 differs from those of other spiral galaxies in the filament, further supporting its identification as a background galaxy. 

\section{Discussion and Conclusion}
\label{sect:discussion}
After excluding galaxies closer to filaments, \citet{2014MNRAS.440L.106A} discovered smaller, filament-like structures they refer to as "tendrils" embedded within the voids \citep{2018ApJ...852..142C}. On average, a tendril contains just under six galaxies and measures approximately 10 Mpc in length. They also suggested that tendrils are morphologically distinct from filaments, appearing more isolated. \citet{2013AJ....145..120B} identified a smaller filament composed of three interacting galaxies from the \HI observation in a void. This cannot rule out the possibility that this triploid is just a  tendrils. Here we discovered a thin filament in the Ursa Major supergroup. We measured the projection length of the filament to be approximately 0.9 Mpc, based on the distance between its two endpoint galaxies. The thin filament contains five gas clumps and sixteen galaxies. Regardless of the number of galaxies it contains, its length, and its position in the cosmic structure, it implies that the filament we discovered is not the tendrils.

Following the multiple lines of evidence, we have detected a coherent galaxy filament with high confidence. The total baryonic mass  of the filament containing all sources is estimated to be 8.8$_{-0.4}^{+0.4}\times10^{10}$ \msol.  We obtained that the $\sigma_{\perp}$ of  the galaxy filament is about 86.6$_{-0.9}^{+0.9}$ \kms. The  $M_{\rm tot}$ of the filament is 3.5$_{-0.2}^{+0.2}\times10^{12}$ \msold, calculated using the length of about 0.9 Mpc. The estimated total mass  of the filament is 3.5$_{-0.2}^{+0.2}\times 10^{12}$ \msold, about 40 times its baryonic mass, indicating that dark matter dominates within the filament. Thus, we have identified a filament of the cosmic web in the Ursa Major supergroup.

Recently, \citet{2024ApJ...976L..18P} discovered five galaxies aligned along a straight line in the plane of the sky. Their calculations indicated that this small group is also dominated by dark matter, consistent with the thin filament we identified. Additionally, large filamentary structures have been uncovered through MeerKAT \HI observations. \citet{2025ApJ...980L...2A} identified a narrow straight filament of galaxies extending approximately 5 Mpc at a redshift of z=0.0365. Meanwhile, \citet{2025A&A...701A..86L} reported a filamentary-like overdensity of \HI galaxies at a redshift of z=0.0395; however, they did not find any dynamic information related to these three filaments. \citet{2025arXiv250813053T} detected a 15 Mpc galaxy filament at redshift z=0.032. Their analysis of 14 \HI-selected galaxies revealed a very thin, elongated structure measuring 1.7 Mpc and indicated that the spin axes of the \HI galaxies are significantly more aligned with the cosmic web filament. For the thin filament with a length of 0.9 Mpc found in the nearby Ursa Major supergroup, this is the first filament dominated by dark matter that not only exhibits galaxy-filament spin alignment but also shows a velocity gradient.

The spin of galaxies is a crucial factor in understanding their morphological diversity in a filament.  Here we use a 2D method to study the galaxy-filament spin alignment. \citet{2020MNRAS.491.2864W} and  \citet{2022MNRAS.513.2168T} suggested that using both 2D and 3D methodologies yields consistent results, but only cause a confusion between filaments and walls. Many simulations suggest that the spin axes of low-mass halos tend to align parallel to filaments, while high-mass halos exhibit an orthogonal alignment \citep{2012MNRAS.427.3320C,2013ApJ...762...72T}. However, our observations did not confirm such a halo mass dependence. Previous studies have indicated that the spins of spiral galaxies generally align parallel to filaments, while S0 galaxies tend to have spins perpendicular to the filament direction \citep{2013MNRAS.428.1827T,2013ApJ...779..160Z,2021MNRAS.504.4626K}.  \citet{2020MNRAS.492..153B} also indicated that the spins of galaxies with various morphological types tend to be aligned with cosmic web filaments. We observed the opposite spin-filament alignment for spiral galaxies, S0 galaxies and other types of galaxies. Moreover, galaxies with the highest baryon mass fraction exhibit alignment with their nearest filament \citep{2022MNRAS.513.2168T}. \citet{2025ApJ...983..100W} suggested that filament spin can affect the galaxy spin-filament correlation. Due to observing a thin filament, we were unable to observe the effect of the rotation on these correlations. Furthermore, numerical simulations have shown that the spin-filament alignment varies with the thickness of the host filament \citep{2014MNRAS.440L..46A,2018MNRAS.481..414G,2021MNRAS.503.2280G}. Across all redshifts, the thin filaments exhibit the highest proportion of halo spins oriented perpendicular to the filament \citep{2018MNRAS.481..414G}. Hence, the disparities are attributed to our study of a thin filament with a thickness comparable to the diameter of a galaxy.

The spins of most galaxies are typically perpendicular to the filament we identified, indicating that these galaxies have a similar formation pathway. When the spin axes of galaxies are perpendicular to the filament, their formation is often attributed to accretion or merging processes along the filament \citep{2018MNRAS.481..414G,2012MNRAS.427.3320C,2020MNRAS.492..153B}. We have observed a significant velocity gradient along the identified galaxy filament. The presentation of the velocity gradients further implies that there used to be a potential accretion cold flow along the filament \citep{2013AJ....145..120B}, providing the necessary gas for the formation and growth of galaxies within the filament \citep{2009MNRAS.397L..64A,2019Sci...366...97U}. However, we do not know the exact orientation of the cold flow within the filament. Moreover, minor mergers can significantly disrupt the spin of galaxies \citep{2010MNRAS.403.1009M}. The observed spin-filament parallel alignment for the S0 galaxy NGC 3931 suggests that its progenitor may have experienced a merging event perpendicular to the spine of the filament.

How did this thin filament become integrated into the supergroup?  According to the prevailing paradigm of structure formation, galaxy clusters and groups are thought to form through a hierarchical sequence of mergers and the accretion of smaller systems \citep{1991ApJ...379...52W,1995MNRAS.275...56N}. The Ursa Major supergroup contains many of which are spiral galaxies. This contrasts with mature clusters like Virgo or Coma, which are dominated by elliptical galaxies. Furthermore, prior research also indicates that the Ursa Major supergroup possesses a low velocity dispersion \citep{1996AJ....112.2471T}, and no X-ray emitting intragroup gas has been detected \citep{2001A&A...370..765V}. These make the Ursa Major supergroup seem like a pre-cluster evolving into a mature galaxy cluster in the context of hierarchical structure formation.  While the identified thin filament, as a smaller group, may have been merged into the Ursa Major supergroup in the hierarchical way. And as for how this thin filament is formed,  more numerical simulations are needed. Next, we will systematically search for thin filaments in the FAST survey data.

\section*{Acknowledgements}

We thank the referee for insightful comments that improved the clarity of this manuscript. 
We acknowledge the support of the National Key R$\&$D Program of China No. 2022YFA1602901. This work is also supported by the National Natural Science Foundation of China (Grant Nos. 12373001, 12225303, 12421003), the Chinese Academy of Sciences Project for Young Scientists in Basic Research, grant no. YSBR-063, the Guizhou Provincial Science and Technology Projects (
Supported by the Guizhou Provincial Science and Technology Projects (No.QKHFQ[2023]003, No.QKHFQ[2024]001, No.QKHPTRC-ZDSYS[2023]003, No.QKHJC-ZK[2025]MS015).


\section*{Data Availability}

The SDSS data are available at the Sloan Digital Sky Survey: \href{https://www.sdss.org/}{https://www.sdss.org/}. The Dark Energy Spectroscopic Instrument (DESI) data is available at \href{http://viewer.legacysurvey.org/}{http://viewer.legacysurvey.org/}. The known galaxy table is available at NASA/IPAC Extragalactic Database: \href{http://ned.ipac.caltech.edu/}{http://ned.ipac.caltech.edu/}. The other data that support the results of this study are available from the corresponding author upon reasonable request.



\bibliographystyle{mnras}
\bibliography{references}

%%%%%%%%%%%%%%%%%%%%%%%%%%%%%%%%%%%%%%%%%%%%%%%%%%

%%%%%%%%%%%%%%%%% APPENDICES %%%%%%%%%%%%%%%%%%%%%

%\appendix


%If you want to present additional material which would interrupt the flow of the main paper, it can be placed in an Appendix which appears after the list of references.

%%%%%%%%%%%%%%%%%%%%%%%%%%%%%%%%%%%%%%%%%%%%%%%%%%

% Don't change these lines
\bsp	% typesetting comment
\label{lastpage}
\end{document}